\pdfoutput=1

\documentclass[
  fontsize=12pt,
  fleqn,
  final,
]{scrartcl}

\RequirePackage{metadata/manuscript}

\begin{document}


\thispagestyle{empty}
{
  \raggedright 
  \Large
  \bfseries
  Four-dimensional topological Yang-Mills-Higgs theories with BRST instability\par 
  \bigskip
  \bigskip
}

\begin{addmargin}[6mm]{0mm}
  {
    \noindent
    \bfseries
    Guilherme~Sadovski
    \bigskip
  }

  {
    \noindent
    \footnotesize
    \rmfamily
    e-mail: \href{mailto:gsadovski@proton.me}{gsadovski@proton.me}
    \bigskip
  }

  {
    \noindent
    \bfseries
    Abstract.\normalfont{} We show that four-dimensional topological Yang-Mills theories, when suitably coupled to Higgs-like fields, admit representations in terms of massive gauge fields in a non-trivial neighborhood of the minima moduli. In the adjoint representation, BRST instability is present beyond tree-level, and closely resembles the Coleman-Weinberg mechanism. The fundamental representation requires realification $SU(N) \hookrightarrow SO(2N)$, but exhibits a more pronounced instability at tree-level. Stable topological solitons (vertices/monopoles) are generically present in the adjoint case. These instabilities indicate the reintroduction of local degrees of freedom into the physical spectrum of these topological field theories. In particular, the realified fundamental case may provide a promising framework for 4d topological gravity models. In addition, our results offer rare examples of BRST symmetry breaking in non-Abelian gauge theories with trivial Gribov problems.
  }
\end{addmargin}

\null{}


\section{Introduction}%
\label{sec:introduction}

Dynamical BRST\footnote{Becchi-Rouet-Stora-Tyutin.} symmetry breaking is an exceptionally rare phenomenon in quantum field theory. This is largely because the nilpotency of the BRST charge $Q$, and its overall cohomological structure, is protected by a topological invariant of the field space --- the Witten index $\tr {(-1)}^F$~\cite{witten1981a,witten1982a,fujikawa1983a}. A non-vanishing index guarantees the balance between bosonic and fermionic zero modes, ensuring that the BRST symmetry remains unbroken. Physical observables can then be consistently defined in the classical and quantum realm through its cohomology groups. Not coincidently, most consistent BRST invariant quantum field theories are defined in the $Q \lvert 0 \rangle = 0$ regime.

Historically, any sign of BRST symmetry breaking was regarded as a pathology --- an indication that the gauge fixing and/or quantization procedure had failed to properly isolate the subspace of physical states. However, the modern understanding has reshaped this perspective. In the Gribov–Zwanziger framework, for instance, the BRST instability --- made explicit by the restriction of gauge field configurations to the Gribov region ---, is interpreted as a possible hint of color confinement in non-Abelian gauge theories~\cite{gribov1978a,zwanziger1989a,maggiore1994a,sorella2009a,dudal2012a,vandersickel2012a}\footnote{Recent developments have shown that the breaking is not catastrophic, and Gribov-Zwanzinger theory can be recast in a new BRST invariant form in several different choices of gauge functions~\cite{pereira2015a,capri2015a,pereira2016a,pereira2016b,pereira2016c}.}. Similarly, in the Parisi–Sourlas framework of stochastic dynamics, unstable BRST symmetries can be related to the emergence of chaos, long-time correlations, $1/f$ noise, \textit{etc.}~\cite{parisi1982a,ovchinnikov2016a,ovchinnikov2025a}. What was once considered a breakdown in consistency, is now understood as a window into non-perturbative phenomena.

In this work, we analyze spontaneous BRST and gauge symmetry breaking in a class of topological quantum field theories (TQFTs) known as \textquote{topological Yang-Mills} (TYM)\footnote{There is no standard use of the term \textquote{TYM} in the literature. It can mean field theories whose Lagrangian density is $\tr (F \wedge F)$, also referred to as Baulieu-Singer theories~\cite{baulieu1988a}. But, it can also mean twisted $\mathcal{N}=2$ Super-Yang-Mills theories, also referred to as Donaldson-Witten theories~\cite{witten1988d,witten1991a}. We adopt the former meaning (a comparison between the two can be found in~\cite{junqueira2021a}).}. As we briefly review in Section~\ref{sec:tym}, these are $SU(N)$ gauge invariant topological field theories, whose BRST quantization give exact path integral representations for the so-called Donaldson polynomials --- smooth invariants of 4-manifolds, carrying global information about differentiable structures~\cite{donaldson1983a,donaldson1990a,donaldson1990b,witten1988d,baulieu1988a,witten1991a}. Since the most current experimental and observational data support the existence of four large physical dimensions~\cite{lee2020a,wenhai2020a,chakravarti2020a,visinelli2018a,vagnozzi2019a}, TYM theories acquire immediate relevance in our understanding of gravity, and the large-scale structure of spacetime~\cite{witten1988a,brans1992,maluga1996,maluga2007}.

TYM theories, remarkably, also have relevance in the non-perturbative sector of quantum chromodynamics (QCD). This is because Donaldson polynomials are intrinsically related to intersection forms on the Yang-Mills (YM) instanton moduli. These are precisely the non-perturbative field configurations responsible for the huge degeneracy of the QCD vacuum. In fact, the instantonic vacuum-to-vacuum amplitudes can change the chirality of fermions, explaining the $U(1)$ axial anomaly and, consequentially, the huge mass of the $\eta '$ meson~\cite{thooft1976a,thooft1986a}. Clearly, it is not hard to image that results in TYM theories can, potentially, trickle down to phenomenological implications in particle physics.

TYM theories are cohomological field theories (Witten-type TQFTs). This means their defining symmetry is a \textquote{topological BRST symmetry}: the BRST operator is the differential for an equivariant cohomology model of a field space. On one hand, this is what enforces the subspaces of physical states to be exclusively populated by BRST invariant vacua ($\exists$ no local excitations), also constraining the set of observables to be purely topological in nature. On the other, it obviously hinders any naïve attempts to directly connect TYM theories with traditional local physics.

Interestingly enough, TYM theories have vanishing Witten index. Its topological BRST symmetry is unprotected, and possibly unstable. This is precisely the focus of our investigation, aiming to construct a bridge between TYM and local physics. Long story short, we have found that, when adequately coupled to \textquote{Higgs-like fields}, these theories do exhibit BRST instability with the emergence of mass scales, revealing non-topological Higgs-like phases. In particular, with all the fields in the adjoint representation of $SU(N)$, BRST is unstable beyond the tree-level --- implying that the effective theories have non-trivial quantum mechanical regimes generically populated with local degrees of freedom: massless and massive gauge field states, and Nambu-Goldstone fermionic field states. There is also the widespread presence of non-local field states associated with stable topological solitons. With the Higgs-like fields  in the fundamental representation, we have found that no spontaneous symmetry breaking occurs unless we appeal to realification. The realified $SU(N) \hookrightarrow SO(2N)$ theories, strikingly, have instability and emergence of a physical mass scale at tree-level. Their classical Higgs-phase dynamics exhibit, explicitly, the presence of the Nambu-Goldstone fermion, while stable topological solitons are generically absent --- just a single vertex/monopole is present in $N=2$. Anyhow, the adjoint case might be an interesting realization of the Coleman-Weinberg mechanism~\cite{coleman1973a}, while the realified fundamental case might inspire more direct connections between high energy topological gravity models and low energy geometrodynamics~\cite{sako1997a,mielke2011a,alexander2016a,sadovski2017a,chengzheng2017a,agrawal2020a,edery2023a,tianyao2023a,kehagias2021a,raitio2024a,sadovski2025a}.

This work is structured as follows. In Section~\ref{sec:tym}, we define our conventions, establish the mathematical framework, and review the main introductory features of TYM theories. The mathematical framework is quite rigorous, but the universal bundle construction provides one of the least convoluted ways to introduce several aspects of these theories. In Section~\ref{sec:adjoint}, we develop the so-called \textquote{Fujikawa-Higgs sector}, with all fields in the adjoint representation of $SU(N)$. We evaluate the Higgs-phase effective theories, argue for the BRST instability beyond tree-level, and describe the effective field states. In Section~\ref{sec:fundamental}, we follow very closely the methodology employed in Section~\ref{sec:adjoint}, but with the \textquote{Fujikawa-Higgs fields} in the fundamental representation. Realification follows, we evaluate the realified Higgs-phase effective dynamics, argue for the tree-level BRST instability, and describe the effective degrees of freedom present. Finally, in Section~\ref{sec:conclusions}, we present our concluding remarks.

\section{Topological Yang-Mills theories}\label{sec:tym}

Let spacetime be $\mathbb{R}^{4}_{\infty}$, endowed with a globally flat metric $\mathrm{g}$. The 4-manifold $\mathbb{R}^{4}_{\infty}$ is the one-point compactification of the standard $\mathbb{R}^4$, compatible with its conformal structure at infinity. This is a convenient choice of background for TYM, because it is compact, Riemannian and 4-dimensional --- necessary conditions for the rigorous definition of the YM instanton\footnote{$\mathbb{R}^{4}_{\infty}$ is also homeomorphic, but not diffeomorphic to $S^4$. Additionally, $\mathrm{g}$ is conformally equivalent to a round metric on the 4-sphere. Thus, CFTs on $\left( \mathbb{R}^{4}_{\infty}, \mathrm{g}\right)$ can pick up non-trivial topological effects while conveniently defined on a flat background.}. Due to the nature of topological field theories, we prefer to work with differential forms. For future reference, $\wedge$ is the exterior product, $d$ is the exterior derivative, $v \rfloor$ is the interior product along the smooth vector field $v$, and $\star$ is the Hodge (star) dual, defined with respect to $\mathrm{g}$. All these operations occur on the exterior algebra $\bigwedge \mathbb{R}^{4}_{\infty}$ of spacetime.

Let $SU(N)$, with Lie algebra $\mathfrak{su}(N)$, be our structure group. This is a physically relevant choice, but also mathematically convenient as $SU(N)$ YM instantons are reasonably well-understood\footnote{The moduli of irreducible $SU(N)$ YM instantons is finite-dimensional, generically smooth and Kähler. Additionally, $SU(N)$ YM instantons can be fully classified via the ADHM construction. The same cannot be said about other choices of symmetry group.}. Also, for future reference, we define on each event $x \in \mathbb{R}^{4}_{\infty}$ the $\mathbb{Z}_2$ graded Lie bracket $[X,Y] \equiv X \wedge Y-{(-1)}^{|X||Y|}Y \wedge X$, where $|X|$ is the grading (statistics) of $X$; the adjoint representation $\mathrm{ad}_X Y \equiv \left[X, Y\right]$, and; the Killing form $\kappa \left(X,Y\right) \equiv \tr \left( \mathrm{ad}_{X} \mathrm{ad}_{Y} \right)$ acting on $X, Y \in \mathfrak{su}(N) \otimes \bigwedge_x \mathbb{R}^{4}_{\infty}$.

The global framework of an $SU(N)$ gauge theory is that of the universal bundle $[ SU(N)\times\mathcal{G} ] \hookrightarrow ( P \times \mathcal{A} ) \rightarrow ( \mathbb{R}^{4}_{\infty} \times \mathcal{M} )$~\cite{baulieu1985a,baulieu1988a}. The smooth  $(N^2+3)$-manifold $P$ is part of the more commonly known principle bundle structure $SU(N) \hookrightarrow P \rightarrow \mathbb{R}^{4}_{\infty}$, together with its adjoint bundles, $ \mathrm{Ad} P \equiv P \times_{SU(N)} SU(N) $ and $ \mathrm{ad}P \equiv P \times_{SU(N)} \mathfrak{su}(N)$. The space of $SU(N)$ connections on $ P $ is $ \mathcal{A} \equiv C^{ \infty } ( J^1 P ) $, where $ J^1 P $ is the 1st jet bundle of $P$, and; the space of smooth gauge transformations is $ \mathcal{G} \equiv C^{ \infty } ( \mathrm{Ad}P ) $, where $ \mathrm{Lie} (\mathcal{G}) \equiv C^{ \infty } ( \mathrm{ad} P ) $ is its Lie algebra. Finally, $ \mathcal{M} \equiv \mathcal{A}/\mathcal{G}$ is the gauge moduli\footnote{The definition of a universal bundle requires a free action of $SU(N) \times \mathcal{G}$ on $P \times \mathcal{A}$. Thus, it is implicit that we are only considering framed irreducible connections, preventing singularities in $\mathcal{M}$ and the ill-posedness of its cohomological problem.}.

A gauge field $A(x)$ is defined as an element in $C^\infty ( \mathrm{ad}P \otimes \bigwedge^1\mathbb{R}^{4}_{\infty} )$, which results from a pullback to spacetime of an $SU(N)$ connection living on $P$. Similarly, a universal gauge field $\tilde{A}(x)$ is defined as the result of a pullback to $\mathbb{R}^{4}_{\infty} \times \mathcal{M}$ of a universal $[SU(N) \times \mathcal{G}]$ connection living on $P \times \mathcal{A}$. They are related by
\begin{equation}
  \label{eq:universal-connection}
  \tilde{A} = A + c \;,
\end{equation}
where $c(x) \in C^\infty ( \mathrm{ad}P \otimes \bigwedge^0\mathbb{R}^{4}_{\infty} )$ is a projection of the Maurer-Cartan form on $ \mathrm{ Lie } (\mathcal{G})$ --- the well-known Faddeev-Popov ghost field. The universal exterior derivative $\tilde{d}$ can also be expressed accordingly,
\begin{equation}%
  \label{eq:universal-exterior-derivative}
  \tilde{d} = d + s \;,
\end{equation}
where $s$ is a projection of the exterior derivative on $ \mathrm{Lie} (\mathcal{G})$ --- the well-known BRST operator.

The $\mathbb{Z}_2$ grading we mentioned earlier is actually a $\mathbb{Z}_2$ bi-grading that exists in the exterior algebra on $P \times \mathcal{A}$, but which descends to spacetime as the differential form rank, and ghost number. Thus, equations~\eqref{eq:universal-connection} and~\eqref{eq:universal-exterior-derivative} is just a bi-grading split into \mkbibquote{components} $(1,0)$ and $(0,1)$. In particular, they give the meaning of an 1-form with ghost number 0 to $A$, and a 0-form with ghost number 1 to $c$. Accordingly, the statistics of $A$, and $c$, is odd (fermionic) --- always check Table~\ref{tab:gradings} for reference. By definition, $\tilde{d}$, $d$, and $s$, are all nilpotent operators, resuming~\eqref{eq:universal-exterior-derivative} to $sd + ds = 0$, and implying their fermionic nature.

The universal curvature $ \tilde{F} $ of $\tilde{A}$ is given by\footnote{From now on, whenever the context is sufficiently clear, we will omit $\wedge$. In particular, $X \wedge X = XX = X^2$.}
\begin{subequations}%
  \label{eq:universal-curvature}
  \begin{align}
    \tilde{F} & \equiv \tilde{d}\tilde{A}+\tilde{A}^2 \;, \\
              & = F+\psi+\phi \;,
  \end{align}
\end{subequations}
and it has curvature $F\equiv dA+A^2$ of $A$ as the $(2,0)$  component; the so-called \emph{topological ghost field} $\psi \equiv sA + Dc$ as the $(1,1)$ component, and; the \emph{2nd generation topological ghost field} $\phi \equiv sc + c^2$ as $(0,2)$ component. We must clarify $D \equiv d+ \left[A, \hphantom{A}\right] $ is the exterior covariant derivative in the adjoint representation.

The Bianchi identities for $\tilde{F}$, and $F$, give
\begin{subequations}%
  \label{eq:universal-bianchi-identity}
  \begin{align}
    \tilde{D}\tilde{F}          & =0 \;,  \\
    sF+D\psi+ \left[c, F\right] & = 0 \;.
  \end{align}
\end{subequations}
And, from them, we can read the so-called \textit{TYM BRST symmetries}~\cite{baulieu1988a},
\begin{subequations}\label{eq:tym-brst}
  \begin{align}
    sA    & = -Dc + \psi \;,                     \label{eq:tym-brst-A}    \\
    sc    & = - c^2 + \phi \;,                   \label{eq:tym-brst-c}    \\
    s\psi & = -D\phi -  \left[c, \psi\right]  \;,\label{eq:tym-brst-psi}  \\
    s\phi & = - \left[ c, \phi \right]  \;,       \label{eq:tym-brst-phi} \\
    sF    & = -D\psi - \left[ c, F \right]  \label{eq:tym-brst-F} \;.
  \end{align}
\end{subequations}
This is the most general realization of the BRST algebra, capturing the full cohomological structure of traditional gauge theories\footnote{It is possible to augment this algebra, and the universal bundle construction, by considering the anti-BRST transformations~\cite{birmingham1991b,carvalho1992a,perry1993a}. However, no extra cohomological information is gained.}. In mathematical language,~\eqref{eq:tym-brst} is called the \emph{BRST/Kalkman model} of the $\mathcal{G}$-equivariant cohomology of $\mathcal{A}$. Imposing $c=0$ in~\eqref{eq:tym-brst-A}, \eqref{eq:tym-brst-psi}, and~\eqref{eq:tym-brst-F}, we get the so-called \emph{Weil model}, and imposing $c=0$ everywhere, we get the \emph{Cartan model} --- Witten's original \textquote{BRST-like supersymmetry}~\cite{witten1988d,ouvry1989a,kanno1989a,kalkman1993a}. Most importantly, these models are isomorphic to each other in the subspace of basic forms\footnote{The definition of a basic form will be more conveniently given in the next Section.}~\cite{baal1990a,witten1991a,cordes1995a}. 

Physically,~\eqref{eq:tym-brst} represents a much stronger set of symmetries than the usual YM BRST from particle physics --- the latter can be obtained from the former by imposing the so-called \emph{horizontal conditions}, $\psi=\phi=0$. Additional two very important aspects of~\eqref{eq:tym-brst} are: (i) as already mentioned in the introduction, its Witten index vanishes --- hinting the possibility of BRST instability~\cite{witten1981a,witten1982a,fujikawa1983a}, and; (ii) also already mentioned, the cohomology groups it defines forbids the presence of $ \mathrm{g} $ metric-contaminated observables, including the traditional YM Lagrangian density $ \tr \left( F \star F \right) $\footnote{Implicitly $\mathrm{g}$ metric-contaminated due to the explicit presence of $\star$.}. Assuming BRST stability, all non-trivial local observables are elements in the $s$-cohomology modulo $d$-boundaries. Therefore, the only ones allowed are
\begin{equation}
  \label{eq:tym-observables}
  \mathcal{O}_n = \tr \left( \tilde{F}^n \right) \; ; \; n \in {\mathbb{N}}^{\ge 1} \;.
\end{equation}
These are invariant polynomials of $\tilde{F}$ which give Chern classes of the universal bundle. In other words, they are all topological in nature. In particular,
\begin{equation}
  \label{eq:donaldson-polynomials}
  \mathcal{O}_2 = \tr \left[ F^2 + 2\psi F + \left( 2\phi F + \psi^2 \right) + 2\psi \phi + \phi^2\right]
\end{equation}
contains precisely the Donaldson polynomials first evaluated in the seminal works of S.~K.~Donaldson~\cite{donaldson1983a,donaldson1990a}, and E.~Witten~\cite{witten1988d} --- albeit for the case $N=2$\footnote{$N>2$ represent generalizations of the original polynomials, reflecting the slightly more involved nature of $SU(N>2)$ bundles over $\mathbb{R}^4_{\infty}$.}.

Among all allowed observables in~\eqref{eq:tym-observables}, only the (4,0) component of~\eqref{eq:donaldson-polynomials} is suitable for a Lagrangian density in dimension four. Thus, the TYM symmetries force the action functional to be
\begin{equation}%
  \label{eq:tym-action}
  S_{\text{TYM}}\left[ A \right]  \equiv  \int \tr\left(F^2\right)
  = 8 \pi^2 k \;,
\end{equation}
where $ k \in \mathbb{Z} $ is the \textit{instanton number}\footnote{The instanton number is generally non-vanishing on $\mathbb{R}^{4}_{\infty}$. It is well-known that $F^2$ is a global $d$-cycle. Poincaré lemma then implies $F^2$ is a local $d$-boundary. In fact, $F^2 = d \omega^{(3)} $, where $\omega^{(3)} \equiv AdA+\tfrac{2}{3}A^3$ is the famous Chern-Simons 3-form. Nonetheless, the 4th de Rham cohomology group of $\mathbb{R}^{4}_{\infty}$, namely $ H^4_{\text{dR}} \left( \mathbb{R}^{4}_{\infty}\right) \sim H^4_{\text{dR}} \left( S^{4}\right) \sim \mathbb{Z} $, obstructs $F^2$ to be a global $d$-boundary, \textit{i.e.}, under an integral sign $F^2$ does not equate $d \omega^{(3)}$:~\eqref{eq:tym-action} does not generically vanish.} of $A$. More rigorously, $k$ is the 2nd Chern number of the $SU(N)$ bundle over $\mathbb{R}^{4}_{\infty}$, on which the principal connection associated with $A$ was originally defined. This is an important clarification because $k$ classifies these bundles. Thus, gauge fields (modulo gauge transformations) are in one-to-one correspondence with them (modulo bundle isomorphisms). In this sense, TYM theories are explicitly topological, and~\eqref{eq:tym-observables},~\eqref{eq:donaldson-polynomials}, and~\eqref{eq:tym-action} are, explicitly, smooth bundle invariants.

Unsurprisingly, the field equations are trivial ($0=0$), signaling lack of local bulk dynamics. Many topological field theories can be formulated as fully extended functorial field theories~\cite{atiyah1988a,segal1988a,baez1995a,schreiber2009a,baez2009a}. In this context, their non-local bulk (or boundary) dynamics can be roughly understood as the propagation and scattering of smoothly embedded fully extended cobordisms. Moreover, TYM theories have vanishing (pure gauge) energy functionals on the bulk, but not necessary on the celestial sphere at infinity
\begin{equation}%
  \label{eq:tym-energy}
  \lim_{ r \to \infty}  \mathbb{E}_{\text{TYM}} [A; r] = \int_{S^{2}_{\infty}} \tr \left[2\left(\lambda \rfloor A\right)F\right] \;,
\end{equation}
where $\lambda$ is the globally constant vector field generating spacetime translations, and $S^{2}_r \equiv \partial B^3_r$ is the 2-sphere boundary of the 3-ball $B^3_r \subset \mathbb{R}^{4}_{\infty} $ of radius $r$. In this sense, every bulk gauge field configuration amounts to the gauge vacua.

Partition functions for TYM theories, formally defined in the weakly coupled regime, require~\eqref{eq:tym-action} to be gauge fixed. Quantum TYM (QTYM) theories have been studied in several different gauge choices~\cite{baulieu1988a,myers1990c,brandhuber1994a,piguet1995a,sadovski2017c,sadovski2018a,sadovski2018b}. A particularly convenient one is the (anti-)self-dual Landau ((A)SDL) conditions 
\begin{subequations}\label{eq:asdlg}
  \begin{align}
    d \star A    & = 0 \;, \\
    d \star \psi & = 0 \;, \\
    F^\pm        & = 0 \,,
  \end{align}
\end{subequations}
where $ F^\pm \equiv F \pm \star F $. This choice localizes the path integral on the very well-behaved $SU(N)$ YM (anti-)instanton moduli, where the Donaldson invariants were originally formulated via intersection theory~\cite{donaldson1983a,donaldson1990a,donaldson1990b}. The (A)SDL gauge results in a very strong set of Ward identities: QTYM theories are renormalizable to all orders in perturbation theory with only one independent, and non-physical, renormalization parameter~\cite{sadovski2017c}, and; all their connected $n$-point Green functions are tree-level exact~\cite{sadovski2018a}. Clearly, this gauge makes evident that QTYM are trivial as local QFTs. And, due to their lack of loop corrections, their quantum observables are the classical observables in~\eqref{eq:tym-observables}: the Donaldson polynomials. In this sense, QTYM and classical TYM represent the same physical theories. 

A final unusual feature worth notice relates to the so-called \textit{Gribov ambiguities}~\cite{gribov1978a}. These ambiguities are present in every non-Abelian gauge theories, including TYM, due to the non-triviality of the gauge moduli bundle $\mathcal{G} \hookrightarrow \mathcal{A} \rightarrow \mathcal{M}$~\cite{singer1978a}. However, different from most non-Abelian gauge theories, TYM theories are tree-level exact. Therefore, a refinement of the gauge fixing procedure \textit{\`a la} Gribov-Zwanzinger, for instance, cannot yield any interesting new physics. In this sense, we say, their Gribov problems are trivial~\cite{sadovski2020a}.

\section{Adjoint TYMH theories}%
\label{sec:adjoint}

We start our investigation on the possible BRST and gauge symmetry breaking, recalling that the traditional Higgs action functional is $\mathrm{g}$ metric-contaminated, and not an $s$-boundary. This is not an ideal candidate to couple to TYM theories, as such coupling would explicitly break the TYM symmetries, while the naïve introduction of a scalar field would give non-vanishing contributions to the Witten index of the surviving BRST symmetry. Thus, the spontaneous symmetry breaking scenario would be completely jeopardized.

To circumvent these issues, we build upon the Higgs-like toy model first proposed by K.~Fujikawa in~\cite{fujikawa1983a}. Fujikawa's original work is not gauge invariant, so what follows is our gauge invariant generalizations of it, first in the adjoint, and later in the fundamental (and realified fundamental) representations of $SU(N)$. Albeit we focus on the coupling to TYM theories, the methodology below can be adapted to several other topological gauge field theories, regardless of their spacetime dimensions. It is also important to stress that the BRST and gauge symmetry breakings are not necessarily related to each other. Each breaking can happen independently, on its own terms --- albeit in the examples below BRST happens to follow gauge symmetry breakings.

All of that said, let us consider the field variables $\Phi (x)$, $\eta (x)$, $\xi (x)$, $B (x) \in C^{\infty} ( \mathrm{ad} P \otimes \bigwedge^0 \mathbb{R}_{\infty}^{4} )$, obeying the algebra
\begin{subequations}%
  \label{eq:afh-brst}
  \begin{align}
    s \Phi & = - \left[c, \Phi\right] + \eta \;, \\
    s \eta & = -\left[c, \eta\right]  \;,        \\
    s \xi  & = - \left[c, \xi\right] + B \;,     \\
    s B    & = -\left[c, B\right]  \;.
  \end{align}
\end{subequations}
We will refer to them as adjoint Fujikawa-Higgs (aFH) fields, and their statistics can be found in Table~\ref{tab:gradings}. Introduced as a pair of $s$-doublets, aFH fields can only build gauge invariant polynomials which are $s$-boundaries, thus living outside the $s$-cohomology groups. This is a general result in BRST algebra, known as the \textit{doublet theorem}. This theorem justifies the form of the action functional~\eqref{eq:afh-action-s-exact}, and also implies that the $s$-cohomology groups are explicitly preserved: no new observables are introduced; especially no $\mathrm{g}$ metric-contaminated ones. In other words, the topological structure of TYM theories is untouched. Moreover, $s$-doublets give vanishing contributions to the Witten index of $s$. In this way, their introduction cannot naïvely change the BRST stability or instability scenarios. 

We naturally choose to minimally couple TYM and aFH fields, defining the Fujikawa-Higgs sector of the $SU(N)$ invariant \textquote{adjoint topological Yang-Mills-Higgs} (aTYMH) theories via the action functional
\begin{subequations}%
  \label{eq:afh-action}
  \begin{align}%
    S_{\text{aFH}} \left[ \Xi \right] & \equiv s \int \tr \left\{D \xi \star D \Phi + \xi \Phi \left[m^2 + g \left(B \Phi - \xi \eta\right)\right] \star \mathds{1} \right\} \;,                      \label{eq:afh-action-s-exact} \\
                                      & = \int \tr \left\{ -DB \star D \Phi - D \xi \star D \eta + \left( \mathrm{ad}_{\Phi}D \xi - \mathrm{ad}_{\xi} D \Phi \right) \star \psi \vphantom{m^2} \right. + \nonumber                  \\
                                      & + \left. \left[ m^2 \left( B \Phi - \xi \eta \right) + g {\left( B \Phi - \xi \eta \right)}^{2} \right] \star \mathds{1}  \right\} \;, \label{eq:afh-action-full}
  \end{align}
\end{subequations}
where $\Xi$ is shorthand for $A, \psi, \Phi, \eta, \xi, B$; $m^2$ and $g$ are mass and coupling parameters, respectively, and; $\star \mathds{1}$ is the unitary volume 4-form on spacetime\footnote{One can see that this functional is agnostic about the spacetime dimension.}. This is not the most general minimally coupled $s$-boundary we could have built out of aFH fields. The addition of more terms, \textit{e.g.}, $s \tr (D \xi \star DB)$, most likely benefits its renormalizability properties, but does not alter its symmetry breaking patterns. Thus, we chose not to add them. Additionally, we purposefully omitted all vertices involving $c$ from~\eqref{eq:afh-action-full}, which is equivalent to work with the already mentioned Cartan model --- isomorphic to the Kalkman model on the subspace of basic forms. It can be shown that the integrand in~\eqref{eq:afh-action-s-exact} is indeed a basic form. For convenience, we will work with the Cartan model whenever we are dealing with basic forms. Explicitly,
\begin{subequations}%
  \label{eq:cartan-model-afh-brst}
  \begin{align}
    s A    & =   \psi \;,   & s \Phi & =   \eta \;, & \hspace{200pt} \\
    s \psi & =  - D\phi \;, & s \eta & = 0 \;,      & \hspace{200pt} \\
    s \phi & = 0 \;,        & s \xi  & =  B  \;,    & \hspace{200pt} \\
    s F    & = -D\psi       & s B    & = 0      \;. & \hspace{200pt}
  \end{align}
\end{subequations}
This $s$, in particular, is nilpotent only up to the infinitesimal gauge transformations generated by $\phi$, namely, $\delta_{\phi} \equiv - s^2$. Now, we can make it clear that the integrand~\eqref{eq:afh-action-s-exact} is basic because it is independent of $\phi$, and invariant under $\delta_{\phi}$. This is the definition of a basic form, and the reason why this $s$ is nilpotent on this subspace. The aFH fields also individually fit this definition, further implying they commute with $\phi$.

We focus on the bosonic potential
\begin{equation}%
  \label{eq:potential}
  V_{\text{aFH}} \left(\Phi, B\right) = \tr \left[m^2 B \Phi + g {( B \Phi )}^2\right]
\end{equation}
with non-trivial local minima
\begin{subequations}%
  \label{eq:minima}
  \begin{align}
    B_0    & \equiv \mathrm{diag} \left(b_1, \ldots, b_N \right) \;,             \\
    \Phi_0 & \equiv \mathrm{diag} \left(\varphi_1, \ldots, \varphi_N \right) \;,
  \end{align}
\end{subequations}
where each pair of eigenvalues $(b_j, \varphi_j)$ must solve
\begin{subequations}%
  \label{eq:eigenvalues-conditions}
  \begin{align}
    m^2 b_j + 2g b_j^2 \varphi_j +\alpha        & = 0 \;,  \\
    m^2 \varphi_j + 2g b_j \varphi_j^2  + \beta & = 0 \;,  \\
    \textstyle\sum_{j} b_j                      & = 0 \;,  \\
    \textstyle \sum_j \varphi_j                 & = 0 \;,  \\
    \Delta_k                                    & > 0  \;.
  \end{align}
\end{subequations}
Einstein sum convention is not assumed in~\eqref{eq:eigenvalues-conditions}, and indexes run as $j \in \left\{ 1, \ldots, N \right\} $, $k \in \left\{ 5, \ldots, 2+2N \right\} $. The parameters $\alpha, \beta \in \mathbb{R}$ are Lagrange multipliers enforcing the traceless condition characteristic of $\mathfrak{su}(N)$ matrices, and $\Delta_k$ is the determinant of the leading $k \times k $ principal minor of the bordered Hessian of $V_{\text{aFH}}$. We stress these are sufficient, but not strictly necessary conditions for local minima. For instance, if $N=2$, then only one independent Casimir operator exists, implying $\tr [ {(B \Phi)}^2 ] \propto { [ \tr ( B \Phi ) ] }^2$. In such case, solving~\eqref{eq:eigenvalues-conditions} directly can be completely avoided by simply \textquote{completing the square} in~\eqref{eq:potential}.

We find that the gauge symmetry breaking pattern is exactly the same as the one in the adjoint Higgs from particle physics. $(B_0, \Phi_0)$ are always valued in the Cartan subalgebra $\bigoplus^{N-1} \mathfrak{u}(1)$ of $\mathfrak{su}(N)$, and have stabilizer $\mathfrak{h}_0 = \left\{ X \in \mathfrak{su}(N) \; ; \; \mathrm{ad}_{B_0} X = 0, \;\mathrm{ad}_{\Phi_0} X = 0\right\} $ in the minima moduli $\mathcal{V}_0$. In the trivial neighborhood, $\mathfrak{h}_0 \sim \mathfrak{su}(N)$. Away from it, gauge symmetry is spontaneously broken, $\mathfrak{h}_0 \subset \mathfrak{su}(N)$. Specifically, in a region where the multiplicity of $(b_j, \varphi_j)$ is $n_j=1 \; \forall \; j$ (non-degeneracy), then $\mathfrak{h}_0 \sim \bigoplus^{N-1} \mathfrak{u}(1)$, implying $N(N - 1)$ broken generators, thus $N(N-1)$ massive gauge field components. In a region where $l \in \left\{1, \ldots, m\right\} $ labels distinct pairs of eigenvalues, and $1 < n_l \leq N \; ; \; \sum_l n_l = N$ (some degeneracy), then $\mathfrak{h}_0 \sim \bigoplus_l \mathfrak{su}(n_l) \oplus \bigoplus^{m-1} \mathfrak{u}(1)$, implying $N^2 - \sum_l n_l^2$ broken generators, thus $N^2 - \sum_l n_l^2$ massive gauge fields components. The region of complete degeneracy ($m=1$) is the trivial neighborhood, implying no broken generators, no massive gauge field components, no gauge symmetry breaking.

The action functional~\eqref{eq:afh-action} assumes the following effective form in the neighborhood of $(B_0, \Phi_0)$,
\begin{subequations}%
  \label{eq:afh-eff-action}
  \begin{align}%
    \bar{S}_{\text{aFH}} \left[ \bar{\Xi} \right] & = \bar{s} \int \tr \left\{ D \xi \star D\varphi + \mathrm{ad}_{\Phi_0} D \xi \star A + \xi ( \Phi_0 + \varphi ) \left[ m^2 + g ( B_0 \Phi_0 \right. + \right.                                                      \nonumber   \\
                                                  & \left. + \left.  B_0 \varphi + b \Phi_0 + b \varphi - \xi \eta ) \vphantom{m^2} \right] \star \mathds{1} \right\}     \;,                \label{eq:afh-eff-action-exact}                                                       \\
                                                  & = \int \tr \left\{ \vphantom{m^2} \mathrm{ad}_{B_0} \mathrm{ad}_{\Phi_0} A \star A + \left( \mathrm{ad}_{B_0} \varphi + \mathrm{ad}_{\Phi_0} b \right) D \star A  + \mathrm{ad}_{\Phi_0} \xi D \star \psi \right. +  \nonumber \\
                                                  & - Db \star D \varphi - D \xi \star D \eta - \left( \mathrm{ad}_{B_0} \mathrm{ad}_{A} \varphi + \mathrm{ad}_{\Phi_0} \mathrm{ad}_{A} b \right) \star A + ( \mathrm{ad}_{\varphi} D \xi \; +  \nonumber                          \\
                                                  & - \mathrm{ad}_{\xi} D \varphi + 2 \mathrm{ad}_{\Phi_0}\mathrm{ad}_{A} \xi ) \star \psi + \left[ ( m^2 + 2g B_0 \Phi_0 ) ( b \varphi - \xi \eta ) + g \left( B_0 \varphi \right. + \right. \nonumber                            \\
                                                  & \left. + \left. \left.  b \Phi_0 + b \varphi - \xi \eta \right)^2 \right] \star \mathds{1} \right\} \;, \label{eq:afh-eff-action-full}                                                                                         
  \end{align}
\end{subequations}
where $\bar{S}_{\text{aFH}} [ \bar{\Xi} ] \equiv S_{\text{aFH}} [A, \psi, \Phi_0 + \varphi, \eta, \xi, B_0 + b]$; $\bar{\Xi}$ is shorthand for $A, \psi, \varphi, \eta, \xi, b$, and; $\bar{s}$ is given by~\eqref{eq:afh-bar-brst}. In particular, its quadratic part
\begin{align}%
  \label{eq:afh-eff-quad-action}
  \bar{S}_{\text{aFH}}^{(2)} \left[ \bar{\Xi} \right] & = \int \tr \left\{ \vphantom{m^2} \mathrm{ad}_{B_0} \mathrm{ad}_{\Phi_0} A \star A + \left( \mathrm{ad}_{B_0} \varphi + \mathrm{ad}_{\Phi_0} b \right) d \star A + \mathrm{ad}_{\Phi_0} \xi d \star \psi \right. +  \nonumber \\
                                                      & - \left. db \star d\varphi - d\xi \star d\eta +\left[ ( m^2 + 2g B_0 \Phi_0 ) \left( b \varphi - \xi \eta \right) + g {\left( B_0 \varphi + b \Phi_0 \right)}^2 \right] \star \mathds{1} \right\}
\end{align}
makes very clear that some of the effective gauge fields have mass defined by the mass matrix $\mathrm{ad}_{B_0}\mathrm{ad}_{\Phi_0}$. This is, of course, a direct reflection of the spontaneous gauge symmetry breaking just described. The terms proportional to $d \star A$ and $d \star \psi$ are removable via gauge fixing. And, the massive bosonic pair $(b, \varphi)$ plays the role similar to the Higgs boson field, with $(\xi,\eta)$ being its massive fermionic partner. Notice that $(\xi,\eta)$ kept its mass because the factor $m^2+2gB_0\Phi_0$ did not identically vanish thanks to the \textquote{adjointness} of our fields --- something what will change in the (realified) fundamental case.

One can easily check that no massless fermions are present in~\eqref{eq:afh-eff-quad-action}, which indicates that the topological BRST symmetry is unbroken at tree-level. The effective BRST
\begin{subequations}%
  \label{eq:afh-bar-brst}
  \begin{align}
    \bar{s} A    & =   \psi \;,   & \bar{s} \varphi & =   \eta \;,    & \hspace{200pt}                            \\
    \bar{s} \psi & =  - D\phi \;, & \bar{s} \eta    & = 0 \;,         & \hspace{200pt}                            \\
    \bar{s} \phi & = 0 \;,        & \bar{s} \xi     & =  B_0 + b  \;, & \hspace{200pt}\label{eq:afh-bar-brst-ssb} \\
    \bar{s} F    & = -D\psi       & \bar{s} b       & = 0      \;,    & \hspace{200pt}
  \end{align}
\end{subequations}
indeed remains nilpotent (up to $\delta_{\phi}$), and isomorphic to the original~\eqref{eq:cartan-model-afh-brst}. And, the effective Higgs-phase action~\eqref{eq:afh-eff-action-full} can indeed be written as the $\bar{s}$-boundary~\eqref{eq:afh-eff-action-exact}, automatically making it a $\bar{s}$-cycle\footnote{The use of the Cartan model is justified because the integrand in~\eqref{eq:afh-eff-action-exact} is basic. This is somewhat surprising given to the explicit presence of $A$ outside $D$. We attribute this to the tree-level stability of the BRST\@. As long as $s(B_0, \Phi_0) = (0,0)$, the basicness of $(B,\Phi)$ gets passed down to $(B_0, \Phi_0)$, and to $(b,\varphi)$, ultimately ensuring the basicness of the integrand (and the nilpotency of $\bar{s}$ on it).}. Clearly, the original topological features are still present at tree-level. Nevertheless, if $ \langle 0 \rvert b \lvert 0 \rangle = 0$, then~\eqref{eq:afh-bar-brst-ssb} implies $\langle 0 \rvert \bar{s} \xi \lvert 0 \rangle = B_0 $. As long as $B_0 \neq 0$, this is a sufficient condition for spontaneous BRST symmetry breaking. In turn, Nambu-Goldstone fermionic states must necessarily be present~\cite{fujikawa1983a,witten1982a,witten1981a,salam1974a}. Since such degrees of freedom are absent at tree-level, they must be generated by loop corrections or non-perturbative effects. Saying it differently, the lack of a Nambu-Goldstone fermion in~\eqref{eq:afh-eff-quad-action} implies that the Higgs-phase aTYMH theories must be quantum mechanically non-trivial. This is an interesting scenario which blurs the line between spontaneous and dynamical symmetry breakings. Ultimately, we can conclude that the $\mathcal{G}$-equivariant cohomology of $\mathcal{A}$ no longer fully captures the complete set of Higgs-phase physical states: we are no longer dealing with cohomological TQFTs.

It is also worth analyzing the energy functional evaluated from~\eqref{eq:afh-action},
\begin{subequations}%
  \label{eq:afh-energy}
  \begin{align}
    \mathbb{E}_{\text{aFH}} \left[ \Xi ; r\right] & =  \int_{B^3_r} \tr \left\{ \mathcal{L}_{\lambda}c \star ( \mathrm{ad}_{\Phi} D \xi - \mathrm{ad}_{\xi} D \Phi) + s \left[ \vphantom{m^2} \mathcal{L}_{\lambda} \Phi \star D \xi + D \xi (\lambda\rfloor \star D \Phi) \right. + \right. \nonumber \\
                                                  & \left. + \left. \mathrm{ad}_\xi \star D \Phi (\lambda \rfloor A) + \xi \Phi [ m^{2} + g ( B \Phi - \xi \eta ) ] ( \lambda \rfloor \star \mathds{1} ) \right] \right\}                                                                              \\
                                                  & =  \int_{B^3_r}\tr \left[ \vphantom{m^2} \mathcal{L}_{\lambda} B \star D \Phi - \mathcal{L}_{\lambda} \Phi \star DB + \mathcal{L}_{\lambda} \eta \star D \xi - \mathcal{L}_{\lambda} \xi \star D \eta \right. + \nonumber                          \\
                                                  & + \left. ( \mathrm{ad}_{\Phi} \mathcal{L}_{\lambda} \xi + \mathrm{ad}_{\xi} \mathcal{L}_{\lambda} \Phi ) \star \psi - (\lambda \rfloor \mathfrak{L}_{\text{aFH}}) \vphantom{m^2} \right] \;,
  \end{align}
\end{subequations}
where $\mathcal{L}_{\lambda} \equiv d \lambda \rfloor + \lambda \rfloor d$ is the Lie derivative along $\lambda$, and; $\mathfrak{L}_{\text{aFH}}$ is the Lagrangian 4-form in~\eqref{eq:afh-action-full}. This is clearly a bulk trivial quantity since the non-exact piece --- the one containing two $c$ vertices --- is removable via gauge fixing\footnote{The canonical energy-momentum 3-form of a non-Abelian gauge theory generally does not originate from a basic form. This makes the Kalkman model unavoidable, resulting in the resurrection of $c$. However, if temporal gauge conditions ($\lambda \rfloor A = \lambda \rfloor \psi =0$) are assumed, then $\mathcal{L}_{\lambda}c$ identically vanishes in the bulk.}. In this sense, the situation around the trivial vacuum is quite similar to pure TYM theories, as every classical field configuration in the bulk is part of the vacua moduli. Unsurprisingly, the same can be said about the effective energy functional $\bar{\mathbb{E}}_{\text{aFH}}$, evaluated from~\eqref{eq:afh-eff-action}. Nonetheless, due to the BRST instability at quantum level, we fully expect $\langle 0 \rvert \bar{\mathbb{E}}_{\text{aFH}} \lvert 0 \rangle $ to be bulk non-trivial. From a perturbative QFT perspective, this scenario very closely resembles the Coleman-Weinberg mechanism~\cite{coleman1973a}. And, exemplifies the non-topological effective features of the Higgs-phase quantum aTYMH theories.

Finally, we close the adjoint case making a few comments about the presence of solitonic degrees of freedom around $(B_0, \Phi_0)$. Recalling the adjoint Yang-Mills-Higgs model from particle physics, the inequivalent gauge vacua form the moduli of flat connections --- homotopically equivalent to a point. Thus, the homotopy of the adjoint Higgs sector alone, here $\pi_2 (SU(N)/{U(1)}^{m-1}) \sim \mathbb{Z}^{m-1}$, is sufficient to determine that $m-1$ independent topological solitons of spatial co-dimension 3 stabilizes in $\mathbb{R}^4_{\infty}$. A Riemannian or Lorentzian structure then implies these are vertices or monopoles, respectively. However, at the tree-level of aTYMH theories, the effective gauge vacua form the entire gauge moduli $\mathcal{A}_\mathcal{H}/\mathcal{H}$ --- most definitely not homotopic to a point. Contributions from all these vacua can potentially destabilize Higgs sector solitons, and \textit{vice versa}. Thus, a careful analysis of the complete moduli $[\mathcal{A}_{\mathcal{H}} \times C^{\infty} (\mathrm{ad}P_{\mathcal{H}})] / \mathcal{H} $ seems to be unavoidable in the classical realm\footnote{This is a very hard problem, beyond the scope of this work.}. Notwithstanding, the quantum realm does suffer from BRST instability, implying $\langle 0 \rvert \bar{\mathbb{E}}_{\text{aFH}} \lvert 0 \rangle \neq 0 $ in the bulk. This latter fact localizes the gauge vacua back onto the moduli of flat connections, ensuring that the Higgs sector vertices/monopoles are quantum mechanically stable, and present in the physical spectrum of these theories.

\begin{table}[htpb]
  \caption{Gradings of all TYM fields, and the representation-independent FH fields.}%
  \label{tab:gradings} \begin{tabular}{cccccccccccc} \toprule
    Field      & $A$ & $F$  & $c$ & $\psi$ & $\phi$ & $\Phi$ & $\eta$ & $\xi$ & $B$  \\
    \midrule
    Form rank  & 1   & 2    & 0   & 1      & 0      & 0      & 0      & 0     & 0    \\
    Ghost no.  & 0   & 0    & 1   & 1      & 2      & 0      & 1      & -1    & 0    \\
    Statistics & odd & even & odd & even   & even   & even   & odd    & odd   & even \\
    \bottomrule
  \end{tabular}
\end{table}

\section{TYMH theories}%
\label{sec:fundamental}

We continue our investigation considering the Fujikawa-Higgs (FH) doublets in the fundamental representation of $SU(N)$. For that, we introduce the associated vector bundle $E \equiv P \times_{SU(N)} \mathbb{C}^N$, its endomorphism bundle $\mathrm{End}(E) \equiv P \times_{SU(N)} \mathrm{End}(\mathbb{C}^N)$, and automorphism bundle $\mathrm{Aut}(E) \equiv P \times_{SU(N)} \mathrm{Aut}(\mathbb{C}^N)$. The group of gauge transformations $\mathcal{G}$ has faithful representation in $\mathrm{Aut}(E)$, and its Lie algebra $\mathrm{Lie(\mathcal{G})}$ has it in $\mathrm{End}(E)$. The TYM fields $(A, c, \psi, \phi)$ are now elements in $C^{\infty}(\mathrm{End}(E) \otimes \bigwedge \mathbb{R}^4_{\infty} )$, while the FH doublets $(\Phi, \eta, \xi, B)$ are in $ C^{\infty}(E \otimes \bigwedge^0 \mathbb{R}^4_{\infty} )$. Again, the gradings can be found in Table~\ref{tab:gradings}.

The BRST algebra is given by
\begin{subequations}%
  \label{eq:ffh-fields}
  \begin{align}
    s \Phi & = - c \Phi + \eta \;, & s \Phi^{\dagger} & = \Phi^{\dagger} c + \eta^{\dagger} \;,            & \hspace{100pt} \\
    s \eta & = - c \eta  \;,       & s \eta^{\dagger} & =  - \eta^{\dagger} c                 \;,          & \hspace{100pt} \\
    s \xi  & = - c \xi + B \;,     & s \xi^{\dagger}  & = - \xi^{\dagger} c + B^{\dagger}              \;, & \hspace{100pt} \\
    s B    & = - c B  \;,          & s B^{\dagger}    & = B^{\dagger} c                  \;,               & \hspace{100pt}
  \end{align}
\end{subequations}
where ${}^{\dagger}$ is defined via the canonical Hermitian metric on $E$. As usual, ${}^{\dagger}$ on $\mathrm{End}(E)$-valued fields means Hermitian adjoint\footnote{In our convention, all TYM fields are anti-Hermitian.}, on $E$-valued fields it means conjugate transpose, and on $\mathbb{C}$-valued fields it means complex conjugate.

In full analogy with the adjoint formulation, we define the Fujikawa-Higgs sector of \textquote{topological Yang-Mills-Higgs} (TYMH) theories via the action functional
\begin{subequations}%
  \label{eq:ffh-action}
  \begin{align}%
    S_{\text{FH}} \left[ \Xi \right] & \equiv s \int \mathfrak{Re}\left\{{(D \xi)}^{\dagger} \star D \Phi + {\xi}^{\dagger} \Phi \left[m^2 + g ( {B}^{\dagger} \Phi - {\xi}^{\dagger} \eta ) \right] \star \mathds{1} \right\} \;,         \\
                                     & = \int \re \left\{ \vphantom{m^2} -{(DB)}^{\dagger} \star D \Phi - {(D \xi)}^{\dagger} \star D \eta + {(D \Phi)}^{\dagger} \star \psi \xi + {(D \xi)}^{\dagger} \star \psi \Phi \right. + \nonumber \\
                                     & + \left. \left[ m^2 ( {B}^{\dagger} \Phi - {\xi}^{\dagger} \eta ) + g {( {B}^{\dagger} \Phi - {\xi}^{\dagger} \eta )}^{2} \right] \star \mathds{1}  \right\} \;,
  \end{align}
\end{subequations}
where $\Xi$ is shorthand for $A, \psi, \Phi, \eta, \xi, B, {\Phi}^{\dagger}, {\eta}^{\dagger}, {\xi}^{\dagger}, {B}^{\dagger}$. However, our analysis is quite short-lived because the resulting bosonic potential,
\begin{equation}%
  \label{eq:ffh-potential}
  V_{\text{FH}} ( \Phi, B, {\Phi}^{\dagger}, {B}^{\dagger} ) = \re \left[m^2 B^{\dagger} \Phi + g {( B^{\dagger} \Phi )}^2\right] \;,
\end{equation}
is not bounded from below (or above, for that matter). Assuming $g \neq 0$, one can easily verify that the only critical point, namely,
\begin{subequations}%
  \label{eq:saddle-pts}
  \begin{align}
    \re ({B}^{\dagger}\Phi) & = -m^2/2g \;, \\
    \im ({B}^{\dagger}\Phi) & = 0 \;,
  \end{align}
\end{subequations}
is a saddle. Thus, there is no spontaneous BRST and/or gauge symmetry breaking since there is no analog of~\eqref{eq:afh-eff-action} and/or~\eqref{eq:afh-bar-brst}. Stated even more clearly, in the fundamental representation of $SU(N)$, and with bosonic potential given by~\eqref{eq:ffh-potential}, TYMH theories do not have effective symmetry broken phases.

\subsection{Realified TYMH theories}%

One possible way to move forward is to consider the realification (decomplexification) of the fundamental representation of $SU(N)$. This is done by applying the realification functor $\mathfrak{i}^*$ on $\mathbb{C}^N$, making it \textquote{forget} its $\mathbb{C}$-linear structure~\cite{kostrikin1989a,budinich1988a}. The result is a real $2N$-dimensional vector space $ \mathbb{C}^N_{\mathbb{R}} \equiv \mathfrak{i}^* (\mathbb{C}^N) $, with a linear complex structure $J$\footnote{In fact, the exists an isomorphism $f: (\mathbb{C}^{N}_{\mathbb{R}}, J) \rightarrow (\mathbb{R}^{2N},J')$ such that $J' \circ f = f \circ J$.}. The functor also realifies the morphisms, $ \mathrm{End} ( {\mathbb{C}}^N_{\mathbb{R}} ) \equiv \mathfrak{i}^* [ \mathrm{End} ( {\mathbb{C}}^N ) ] $, $ \mathrm{Aut} ( {\mathbb{C}}^N_{\mathbb{R}} ) \equiv \mathfrak{i}^* [ \mathrm{Aut}( {\mathbb{C}}^N ) ] $, and implies the canonical\footnote{Up to conjugacy in $SO(2N)$.} embedding $ \mathfrak{i}: SU(N) \hookrightarrow SO(2N)$. In practice, if $z \in \mathbb{C}^N$, then $z_{\mathbb{R}} = {[ \re (z), \im (z) ]}^T \in \mathbb{C}^N_\mathbb{R}$, and if $Z \in \mathrm{End}(\mathbb{C}^N)$, then
\begin{equation}%
  \label{eq:realified-matrix}
  Z_{\mathbb{R}} = \begin{pmatrix} \re (Z) & -\im (Z) \\ \im (Z) & \re (Z) \end{pmatrix} \in \mathrm{End}(\mathbb{C}^N_\mathbb{R}) \;.
\end{equation}
The particular form of~\eqref{eq:realified-matrix} ensures its commutativity with $J$ --- a sufficient condition for $Z_{\mathbb{R}}$ to indeed be the realification of $Z$.

From now one, we will consider the realified bundle $E_{\mathbb{R}} \equiv P \times_{SU(N)} \mathbb{C}^{N}_\mathbb{R}$. The TYM fields are realified, $(A_{\mathbb{R}}, c_{\mathbb{R}}, \psi_{\mathbb{R}}, \phi_{\mathbb{R}}) \in C^{\infty}(\mathrm{End}(E_{\mathbb{R}}) \otimes \bigwedge \mathbb{R}^4_{\infty} )$, and so are the FH fields $(\Phi_{\mathbb{R}}, \eta_{\mathbb{R}}, \xi_{\mathbb{R}}, B_{\mathbb{R}}) \in C^{\infty}(E_{\mathbb{R}} \otimes \bigwedge^0 \mathbb{R}^4_{\infty} )$. We stress that realified $SU(N) \hookrightarrow SO(2N)$ topological Yang-Mills (TYM$_{\mathbb{R}}$) theories are not the same as $SO(2N)$ TYM theories. The latter have much bigger moduli than the former, reflecting the more general nature of $SO(2N)$ bundles over $\mathbb{R}^4_{\infty}$. In fact, TYM$_{\mathbb{R}}$ theories only account for those bundles with 1st Pontryagin number $p_1 (E_{\mathbb{R}}) = -2 k$. Accordingly, their action functional is
\begin{equation}%
  \label{eq:real-tym-action}
  S_{\text{TYM}_{\mathbb{R}}} [A_{\mathbb{R}}] \equiv \int \tr_{\mathbb{R}} (F_{\mathbb{R}}^2) = - 8 \pi^2 p_1 \;,
\end{equation}
where $\tr_{\mathbb{R}} (\hphantom{F}) \equiv 2\re [ \tr (\hphantom{F}) ] $ is the realified trace.

The realified BRST algebra remains (mostly) the same\footnote{From now on, we omit the realified label \textquote{$_{\mathbb{R}}$} whenever there is a risk of cluttering equations.},
\begin{subequations}%
  \label{eq:real-brst}
  \begin{align}
    s A    & = - Dc + \psi \;,         & s \Phi & = - c \Phi + \eta \;, & s \Phi^{T} & = \Phi^{T} c + \eta^{T} \;,            & \hspace{0pt} \\
    s c    & = - c^2 + \phi  \;,       & s \eta & = - c \eta  \;,       & s \eta^{T} & =  - \eta^{T} c                 \;,    & \hspace{0pt} \\
    s \psi & = -D \phi - [c, \phi] \;, & s \xi  & = - c \xi + B \;,     & s \xi^{T}  & = - \xi^{T} c + B^{T}              \;, & \hspace{0pt} \\
    s \phi & = - [c, \phi]  \;,        & s B    & = - c B  \;,          & s B^{T}    & = B^{T} c                  \;,         & \hspace{0pt}
  \end{align}
\end{subequations}
where ${}^{T}$ is defined via the Euclidean metric on $E_{\mathbb{R}}$. As usual, ${}^{T}$ on $\mathrm{End}(E_{\mathbb{R}})$-valued fields means adjoint\footnote{In our conventions, all TYM$_{\mathbb{R}}$ fields are antisymmetric.}, on $E_{\mathbb{R}}$-valued fields it means transpose, and on $\mathbb{R}$-valued fields it means just the identity map.

The realified Fujikawa-Higgs (FH$_{\mathbb{R}}$) sector is given by the action
\begin{subequations}%
  \label{eq:real-ffh-action}
  \begin{align}%
    S_{\text{FH}_{\mathbb{R}}} \left[ \Xi_{\mathbb{R}} \right] & = s \int \left\{ {(D \xi)}^{T} \star D \Phi + {\xi}^{T} \Phi \left[m^2 + g ( {B}^{T} \Phi - {\xi}^{T} \eta ) \right] \star \mathds{1} \right\} \;,                      \\
                                                               & = \int \left\{ \vphantom{m^2} -{(DB)}^{T} \star D \Phi - {(D \xi)}^{T} \star D \eta + {(D \Phi)}^{T} \star \psi \xi + {(D \xi)}^{T} \star \psi \Phi \right. + \nonumber \\
                                                               & + \left. \left[ m^2 ( {B}^{T} \Phi - {\xi}^{T} \eta ) + g {( {B}^{T} \Phi - {\xi}^{T} \eta )}^{2} \right] \star \mathds{1}  \right\} \;,
  \end{align}
\end{subequations}
where $\Xi_{\mathbb{R}}$ is shorthand for $A, \psi, \Phi, \eta, \xi, \Phi^T, \eta^T, \xi^T, B^T $, and has bosonic potential
\begin{equation}%
  \label{eq:real-ffh-potential}
  V_{\text{FH}_{\mathbb{R}}} ( \Phi, B^T) = m^2 B^{T} \Phi + g {( B^{T} \Phi )}^2 \;,
\end{equation}
now, bounded from below by the non-trivial minima
\begin{subequations}%
  \label{eq:real-ffh-minima}
  \begin{align}
    {B}_0    & = {(b_0, 0, \ldots, 0)}^{T}  \;,       \\
    {\Phi}_0 & = {(\varphi_0, 0, \ldots, 0)}^{T}  \;,
  \end{align}
\end{subequations}
where $b_0 \varphi_0 = - m^2/2g \; \forall \; g>0$.

We find that the gauge symmetry breaking pattern differs from usual Higgs model from particle physics due to realification. Recalling the standard Higgs, with non-trivial vacua $\Phi_0$, and stabilizer $\mathfrak{h}_0 = \left\{ X \in \mathfrak{su}(N) \; ; \; X \Phi_0 = 0\right\} $ in $ \mathcal{V}_0 $; the condition $X \Phi_0 = 0$ forces the 1st row, and 1st column of $X$ to vanish, making it effectively of rank $N-1$. This is the usual symmetry breaking pattern: in the trivial vacuum neighborhood, $\mathfrak{h}_0 \sim \mathfrak{su}(N)$, while in the neighborhood of $\Phi_0$, we have $\mathfrak{h}_0 \sim \mathfrak{su}(N-1)$. Consequentially, $2N-1$ generators are broken, resulting in $2N-1$ massive gauge field components. Turning to realified Higgs, $\Phi_0$ has stabilizer $\mathfrak{h}_0 = \left\{ X \in \mathfrak{su}(N) \hookrightarrow \mathfrak{so}(2N) \; ; \; X \Phi_0 = 0 \right\}$, where $X$ must have the form~\eqref{eq:realified-matrix}. Notice that if its 1st row, and 1st column vanish, so does its $(N+1)$-th row, and $(N+1)$-th column, making it effectively of rank $2N-2$. Therefore, realification changes the gauge symmetry breaking pattern from $\mathfrak{su}(N) \mapsto \mathfrak{su}(N-1)$ to $\mathfrak{so}(2N) \mapsto \mathfrak{so}(2N-2)$. Since $4N-3$ generators are broken, $4N-3$ gauge field components are massive. Interestingly, realification amplifies\footnote{Roughly analogous to how an embedding $S^{1} \hookrightarrow \mathbb{R}^3$ gives the circle freedom to knot itself, a realification embedding $SU(N) \hookrightarrow SO(2N)$ gives Higgs theories more gauge directions to break.} the gauge symmetry breaking, increasing the number of massive gauge fields components by $2N-2$.

In TYMH$_{\mathbb{R}}$ theories, the neighborhood of the non-trivial vacuum $(B_0,\Phi_0)$ has stabilizer $\mathfrak{h}_0 = \left\{ X \in \mathfrak{su}(N) \hookrightarrow \mathfrak{so}(2N) \; ; \; X B_0 = 0, \; X \Phi_0 = 0 \right\} $, where $X$ has the form~\eqref{eq:realified-matrix}. We thus conclude their gauge symmetry breaking pattern is $\mathfrak{so}(2N) \mapsto \mathfrak{so}(2N-2)$. Consequentially, they admit massive representation around $(B_{0}, \Phi_{0})$ accounting for these $4N-3$ broken generators. Namely,
\begin{align}
  \label{eq:real-eff-action}
  \bar{S}_{\text{FH}_\mathbb{R}} \left[ {\bar{\Xi}}_{\mathbb{R}} \right] & = \int \left[ \vphantom{m^2} {B_0}^T A \star A \Phi_0 - {B_0}^T D \star A \varphi - {\Phi_0}^{T} D \star A b + {\Phi_0}^{T} D \star \psi \xi \right. +  \nonumber \\
                                                                         & - {(Db)}^{T} \star D \varphi - {(D \xi)}^{T} \star D \eta + {B_0}^{T} A \star A \varphi + {\Phi_0}^{T} A \star A b  \; +  \nonumber                               \\
                                                                         & + {(D \varphi)}^{T} \star \psi \xi + {(D \xi)}^{T} \star \psi \varphi - 2 {\Phi_0}^{T} A \star \psi \xi \; + \nonumber                                            \\
                                                                         & + \left. g {( {B_0}^{T} \varphi + b^T \Phi_0 + b^T \varphi - \xi^T \eta )}^2 \star \mathds{1} \right] \;,
\end{align}
with quadratic part given by
\begin{align}%
  \label{eq:real-eff-quad-action}
  \bar{S}^{(2)}_{\text{FH}_\mathbb{R}} \left[ {\bar{\Xi}}_{\mathbb{R}} \right] & = \int \left[ \vphantom{m^2} {B_0}^T A \star A \Phi_0 - {B_0}^T d \star A \varphi - {\Phi_0}^{T} d \star A b + {\Phi_0}^{T} d \star \psi \xi \right. +  \nonumber \\
                                                                               & - \left. d b^T \star d \varphi - d \xi^T \star d \eta +  g {( {B_0}^{T} \varphi + b^T \Phi_0 )}^2 \star \mathds{1} \right] \;,
\end{align}
where $\bar{S}_{\text{FH}_\mathbb{R}} [ {\bar{\Xi}}_{\mathbb{R}} ] \equiv S_{\text{FH}_\mathbb{R}} [ A, \psi, \Phi_0 + \varphi, \eta, \xi, b, \Phi^T_0 + \varphi^T, \eta^T, \xi^T, B^T_0 + b^T] $, and; ${\bar{\Xi}}_{\mathbb{R}}$ is shorthand for $A, \psi, \varphi, \eta, \xi, b, \varphi^T, \eta^T, \xi^T, b^T$.

Equations~\eqref{eq:real-eff-action} and~\eqref{eq:real-eff-quad-action} should immediately be contrasted with their analog~\eqref{eq:afh-eff-action} and~\eqref{eq:afh-eff-quad-action}, respectively, from the adjoint formulation. Among the similarities and differences, we notice the glaring absence in~\eqref{eq:real-eff-quad-action}, of the term proportional to $m^2 + 2g {B_0}^{T} \Phi_0$ --- which identically vanished given to the (realified) \textquote{fundamentalness} of our fields. As a result, $(\xi, \eta)$ is now massless while $(b,\phi)$ remains massive. This implies that their doublet structure is no longer realized in this region of the moduli. Dramatically, the BRST instability of TYMH$_{\mathbb{R}}$ theories is a tree-level phenomenon, and $(\xi, \eta)$ is nothing but the resulting tree-level Nambu-Goldstone fermion.

The effective BRST around $(B_0, \Phi_0)$, in the presence of tree-level instability, assumes the form
\begin{subequations}%
  \label{eq:real-eff-brst}
  \begin{align}
    \bar{s} A    & = \psi \;,    & \bar{s} \varphi & = \eta - \bar{s} \Phi_0\;, & \hspace{200pt} \\
    \bar{s} c    & = \phi  \;,   & \bar{s} \eta    & = 0 \;,                    & \hspace{200pt} \\
    \bar{s} \psi & = -D \phi \;, & \bar{s} \xi     & = b + B_0 \;,              & \hspace{200pt} \\
    \bar{s} \phi & = 0  \;,      & \bar{s} b       & = - \bar{s} B_0  \;.       & \hspace{200pt}
  \end{align}
\end{subequations}
In comparison to its adjoint analog~\eqref{eq:afh-bar-brst}, it has acquired shifts which indeed spoil the doublet and basic nature of the (effective) FH fields. It remains nilpotent (up to $\delta_{\phi}$), but it is no longer isomorphic to~\eqref{eq:real-brst} in the subspace of basic forms. In other words,~\eqref{eq:real-eff-brst} does not represent the Cartan model of the $\mathcal{G}$-equivariant cohomology of $\mathcal{A}$. Even more dramatically,~\eqref{eq:real-eff-action} is not a $\bar{s}$-cycle, automatically implying it also not a $\bar{s}$-boundary.

The Higgs-phase effective theories defined by~\eqref{eq:real-eff-action} have lost every trace of topological BRST invariance --- the non-topological features hidden in the quantum regime of aTYMH theories, are here exposed at tree-level. In particular, the dodge of the doublet theorem by the FH fields, and the presence of $\mathrm{g}$ metric-contaminated terms in~\eqref{eq:real-eff-action}, together imply local (non-topological) degrees of freedom are now present and physically relevant in the bulk. In other words, in this region of the moduli, $\text{TYMH}_{\mathbb{R}}$ theories are clearly standard local field theories.

It is straightforward to verify that the energy functional $\mathbb{E}_{\text{TYMH}_{\mathbb{R}}}$, evaluated from~\eqref{eq:real-ffh-action} is bulk trivial, but $\bar{\mathbb{E}}_{\text{TYMH}_{\mathbb{R}}}$, evaluated from~\eqref{eq:real-eff-action}, is not. Again, due to the tree-level instability. Thus, the effective gauge vacua are already localized onto the moduli of flat connections. Despite this, solitonic degrees of freedom remain quite rare occurrences. The reason is that the quotients $SO(2N)/SO(2N-2)$ are diffeomorphic to Stiefel manifolds $V_2 (\mathbb{R}^{2N})$. Their non-trivial homotopy, $\pi_{2N-2} [ V_2 ( \mathbb{R}^{2N})] \sim \mathbb{Z} $, implies that only for $N=2$, we have a single topological soliton of spatial co-dimension 3 stabilizing in ${\mathbb{R}}_{\infty}^{4}$. A Riemannian or Lorentzian structure then manifests it as a vertex or monopole, respectively.

\section{Conclusions}%
\label{sec:conclusions}

Our analysis have demonstrated that four-dimensional $SU(N)$ invariant TYM theories, when minimally coupled to Fujikawa-Higgs fields, accept representation in terms of massive gauge fields in a non-trivial neighborhood of their moduli. This is a general feature when all fields are in the adjoint representation (aTYMH theories), while it relies on the realification $SU(N) \hookrightarrow SO(2N)$ when the Fujikawa-Higgs fields are in the fundamental representation of the gauge group (TYMH$_{\mathbb{R}}$ theories). In particular, this is a surprising result in the context of topological field theories since they are supposed to be scale invariant theories.

The emergence of mass scales may be physically insignificant if path integrals remain unaffected by said parameters. This would be a scenario similar to how cohomological TQFTs retain scale invariance even when faced with non-vanishing $\beta$-functions of their coupling parameters. Most notably, Witten's $SU(2)$ $\mathcal{N}=2$ twisted Super-Yang-Mills theory~\cite{witten1988d,brooks1988a}. However, for such a scale invariance to be realized, the quantum theories need very strong sets of Ward identities. Decisively, the Slavnov-Taylor identities inherited from topological BRST symmetries. Saying it differently, the theories need topological BRST stability --- precisely what aTYMH and TYMH$_{\mathbb{R}}$ lack in their Higgs phases.

We have shown that the BRST instability of aTYMH theories happens beyond tree-level, and closely resembles the Coleman-Weinberg mechanism if loop corrections are to blame. Accordingly, the emergence of  $N^2 - \sum_l n_l^2$ massive gauge fields only carries physical weight in the quantum realm. Similarly, the tree-level Higgs-phase energy functional $\bar{\mathbb{E}}_{\text{aTYMH}}$ is bulk trivial, but $\langle 0 \rvert \bar{\mathbb{E}}_{\text{aTYMH}} \lvert 0 \rangle$ is not --- hinting a quantum localization of the effective gauge vacua onto the moduli of flat connections. As collateral result, $m-1$ independent topological solitons of spatial co-dimension 3, associated with the spontaneous gauge symmetry breaking $\mathfrak{su}(N) \mapsto \bigoplus_l \mathfrak{su}(n_l) \oplus \bigoplus^{m-1} \mathfrak{u}(1)$, are quantum mechanically stable. We conclude that Higgs-phase quantum aTYMH theories are generically populated with massless and massive gauge field states, Higgs field states, Nambu-Goldstone fermionic field states, as well as vertex or monopole non-local states depending on whether the metric structure of spacetime is Riemannian or Lorentzian, respectively.

Unexpectedly, the BRST instability of TYMH$_{\mathbb{R}}$ theories more pronouncedly happens at tree-level. We can explicitly observe the emergence of a Nambu-Goldstone fermion in the effective action~\eqref{eq:real-eff-action}. We can also observe $4N-3$ massive gauge field components, associated with the realified spontaneous gauge symmetry breaking $\mathfrak{so}(2N) \mapsto \mathfrak{so}(2N-2)$, implying the emergence of a physical mass scale at classical level. Similarly, the tree-level Higgs-phase energy functional $\bar{\mathbb{E}}_{\text{TYMH}_{\mathbb{R}}}$ is bulk non-trivial --- hinting a classical localization of the effective gauge vacua on the moduli of flat connections. Stable solitonic degrees of freedom are, however, quite rare. Only $N=2$ produces a single topological soliton of spatial co-dimension 3. Upon quantization, the Higgs-phase effective TYMH$_{\mathbb{R}}$ theories are generically populated with massless and massive gauge field states, Higgs field states, and Nambu-Goldstone fermionic field states.

Ultimately, topological BRST instability implies that local degrees of freedom are generically present in the Higgs phases of these topological field theories. We speculate that the adjoint case might be an interesting realization of the Coleman-Weinberg mechanism~\cite{coleman1973a}, while the realified fundamental case might bridge high energy topological gravity models to low energy geometrodynamics~\cite{sako1997a,mielke2011a,alexander2016a,sadovski2017a,chengzheng2017a,agrawal2020a,edery2023a,tianyao2023a,kehagias2021a,raitio2024a,sadovski2025a}.

\printbibliography{}

\end{document}